\documentclass[11pt]{article}
\usepackage[a4paper,left=2.4cm,right=2.4cm,top=2.5cm,bottom=2.5cm]{geometry}
\usepackage[utf8]{inputenc}

\usepackage{amssymb,adjustbox,mathtools,amsmath,amsfonts,booktabs,eurosym,fancyhdr,graphicx,caption,color,etoolbox,setspace,sectsty,comment,pdflscape,pdfpages,subcaption,array,float,textcomp,booktabs,gensymb,multirow,url,ragged2e,braket,lipsum,upgreek,titlesec,cite}

\usepackage[hidelinks]{hyperref}

\titlespacing*{\section}
{0pt}{5.5ex plus 1ex minus .2ex}{4.3ex plus .2ex}
\titlespacing*{\subsection}
{0pt}{5.5ex plus 1ex minus .2ex}{4.3ex plus .2ex}

\makeatletter
\renewcommand{\maketitle}{\bgroup\setlength{\parindent}{0pt}
	\begin{center}
		\textbf{\Large\@title}\\
		\vspace{5mm}
		
		\@author
	\end{center}\egroup
}
\makeatother

\date{}
\author{Harold J.W. Zandvliet\textsuperscript{1*},
	Jort D. Verbakel$^{1\dagger}$,
	Qirong Yao\textsuperscript{2},
	Kai Sotthewes\textsuperscript{1} and
	Pantelis Bampoulis\textsuperscript{1}}

\title{\textbf{ 
		\flushleft{Topologically protected one-dimensional electronic states in group IV two-dimensional Dirac materials}
	}
	}
\date{}

\begin{document}

\flushleft

\centering
\maketitle

\begin{center}
	\vspace{1cm}
	{\bf 1} Physics of Interfaces and Nanomaterials, MESA+ Institute for Nanotechnology, University of Twente, P.O. Box 217, 7500AE Enschede, The Netherlands
	\\
	{\bf 2} Center for Artificial Low Dimensional Electronic Systems, Institute for Basic Science (IBS), Pohang 37673, Republic of Korea
	\\
	
	* h.j.w.zandvliet@utwente.nl, $^\dagger$j.d.verbakel@utwente.nl\\ 
	\date{\today}
\end{center}

\vspace{20mm}
\subsection*{Abstract}
\justify
In this report we give a brief introduction on the occurrence of topologically protected one-dimensional electronic states in group IV two-dimensional graphene-like materials. We discuss the effect of spin-orbit coupling on the electronic band structure and show that these materials are potential candidates to exhibit the quantum spin Hall effect. The quantum spin Hall effect is characterized by a gapped interior and metallic counter-propagating spin-polarized topologically protected edges states. We also elaborate on the electric-field induced formation of a hexagonal network of one-dimensional topologically protected electronic states in small-angle twisted bilayer graphene. 

\vspace{10pt}
\noindent\rule{\textwidth}{1pt}
\tableofcontents\thispagestyle{fancy}
\noindent\rule{\textwidth}{1pt}
\vspace{10pt}

\newpage
\setlength{\parindent}{0pt}
\pagestyle{fancy}
\section{Introduction}
\label{sec:intro}
Nowadays we are surrounded by electronic devices, such as smart phones, laptops, tablets, smartwatches, and personal computers. The elementary building block of all these electronic devices is the transistor. The importance of the transistor was recognized soon after its discovery in 1947 by William Shockley, John Bardeen, and Walter Brattain. Owing to the importance of the transistor Shockley, Bardeen, and Brattain received in 1956 the Nobel Prize in Physics "for their researches on semiconductors and their discovery of the transistor effect." The first transistor was made of germanium because of its superior charge carrier mobilities. Germanium was, however, rapidly replaced by silicon. Silicon has a number of advantages over germanium: silicon is much cheaper, has a larger bandgap, is more abundant, and silicon transistors have a longer lifetime and can withstand much higher temperatures. Since its discovery the dimensions and the price of a transistor have exhibited a dramatic decrease, leading to the current generation of electronics containing more than a billion transistors. The costs to produce highly reliable nm-sized transistor is currently less than the prize of a grain of rice. Given the importance of silicon-based transistors for our society the late 20\textsuperscript{th} century and early 21\textsuperscript{st} century are sometimes referred to as the “Silicon Age”.\\
One of the downsides of the ever-expanding silicon-based technology is the continuously growing energy consumption. By 2030 the total energy demand for data storage, wireless and wired networks, production of ICT equipment and electronic consumer devices, such as computers, laptops, smartphones and tablets is expected to be $\sim$20\% of our total global energy consumption \cite{Jones2018}. It is crystal clear this situation is not sustainable and therefore energy minimization routes have to be explored. As most of the energy consumption is due to Joule heating, it would be very beneficial to make use of ballistic transport channels that are completely free of scattering centers. For a defect-free atomic metallic wire, where the transport is carried by a single mode, the resistance is independent of the length of the wire and given by h/2e\textsuperscript{2}. This resistance is not the resistance of the metallic wire itself, but rather the contact resistance between the metallic wire and the electrode. Unfortunately, it is very difficult to prepare nanowires that are completely free of scattering centers and therefore this route requires superb preparation methods. Another route would be to make use of topologically protected one-dimensional channels in which backscattering is forbidden or strongly suppressed because of topological reasons. The edge or boundary modes of two-dimensional topological insulators are protected by symmetries of the nontrivial bulk states. In the case these symmetries are not broken, they can result in novel phenomena, such as the quantum spin Hall effect. In the one-dimensional spin-polarized topologically protected edge states of a quantum spin Hall insulator backscattering is suppressed by time-reversal symmetry $(E(k,\uparrow)=E(-k,\downarrow))$. Backscattering, which changes the sign of the momentum, also requires a flip of the spin. As the spin is preserved for nonmagnetic impurities or defects the backscattering process is forbidden, leading to robust and dissipation-less electronic transport channels along the edges of the material. Another example is small angle twisted bilayer graphene. Upon the application of a transversal electric field a bandgap opens up in the graphene bilayer. This bandgap is inverted for AB and BA stacked regions resulting in a two-dimensional hexagonal network of topologically protected one-dimensional electronic states. This network of one-dimensional states is topologically protected as long as intervalley scattering, i.e. scattering from valley to valley, does not occur.\\
It is the aim of this report to give a brief review on the occurrence of topologically protected states in group IV two-dimensional Dirac materials and stacks thereof. As there are many different types of two-dimensional materials an extensive discussion of the structural and electronic properties of all types of two-dimensional materials goes beyond the scope of this manuscript and therefore we will restrict ourselves to the low atomic number group IV 2D Xenes, i.e. graphene and its silicon, germanium and tin analogues of graphene, referred to as silicene, germanene and tinene or stanene, respectively. We will start off with a brief introduction on the structural and electronic properties of graphene. Since silicene, germanene and stanene are very similar to graphene most, but not all, of the results obtained for graphene are also valid for the other group IV 2D Dirac materials. We will proceed with a brief introduction of the quantum Hall effect and the anomalous quantum Hall effect. Subsequently, we will elaborate on the spin-orbit coupling in two-dimensional graphene-like Dirac materials. We will show, that owing to the opening of a spin-orbit bandgap, graphene and the other group IV 2D Dirac materials are two-dimensional $Z_2$ invariant topological insulators that are expected to exhibit the quantum spin Hall effect. In the next section we will discuss the effect of an applied transversal electric field on the electronic band structure of two-dimensional Dirac materials. For the buckled honeycomb lattices of the group IV 2D Dirac materials, this transversal electric field allows to tune the bandgap. For small electric fields the group IV 2D Dirac materials remain topological insulators, but for transversal electric fields exceeding a critical value these materials undergo a topological phase transition to a normal band insulator. In the next section, we will extend our discussion to bilayers of the group IV two-dimensional Dirac materials. Depending on the exact stacking geometry the lattice symmetry can be broken. For an AB stacked bilayer, the application of a transversal electric field results in the opening of a bandgap. If the two layers are twisted with respect to each other, a moiré structure will develop which consists of an ordered arrangement of AA, AB and BA stacked domains. The electric field induced bandgaps in the AB and BA regions are inverted, resulting in a set of four topologically protected one-dimensional states (two per valley) at the boundaries between the AB and BA domains. 

\section{Electronic structure}

\subsection{Electronic band structure of graphene and the other group IV 2D Xenes}

\begin{figure}[H]
	\centering
	\includegraphics[trim=6cm 6cm 6cm 4cm,clip=true,width=\textwidth]{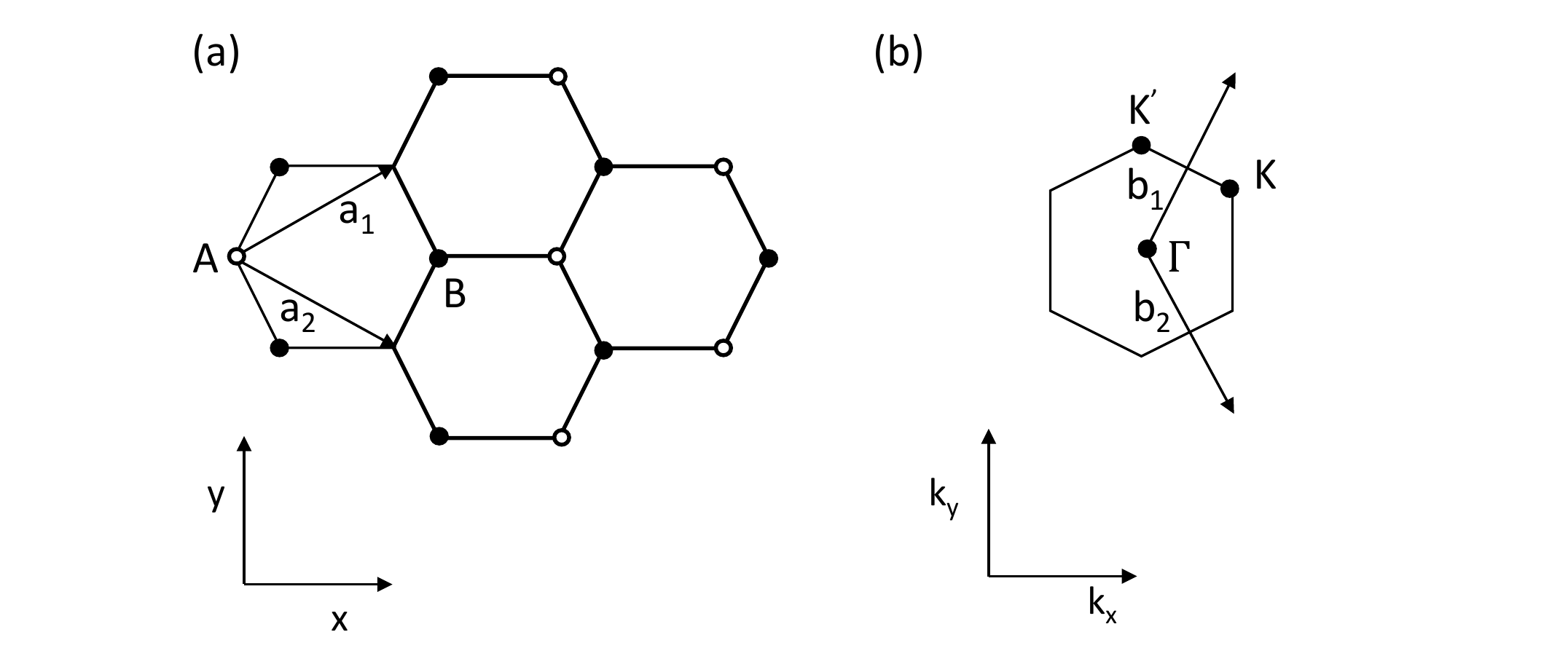}
	\caption{(a) The honeycomb lattice of graphene. $a_1$ and $a_2$ are the real-space lattice vectors. (b) The reciprocal lattice of graphene. $b_1$ and $b_2$ are the reciprocal lattice vectors. $a_1 = \frac{a\sqrt{3}}{2}\mathbf{e_x} + \frac{a}{2}\mathbf{e_y}$, $a_2 = \frac{a\sqrt{3}}{2}\mathbf{e_x} - \frac{a}{2}\mathbf{e_y}$ , $b_1 = \frac{2\pi}{a\sqrt{3}}\mathbf{e_x} + \frac{2\pi}{a}\mathbf{e_y}$ , 
		$b_2 = \frac{2\pi}{a\sqrt{3}}\mathbf{e_x} - \frac{2\pi}{a}\mathbf{e_y}$}
		\label{fig:honeycomb}
\end{figure}

In 2004 graphene, the first two-dimensional material, was successfully isolated by Geim and Novoselov \cite{Novoselov2004,Geim2007}. Graphene is a single layer of carbon atoms that are arranged in a honeycomb structure. The unit cell of the honeycomb lattice involves two carbon atoms, which are labelled A and B, respectively (see Figure \ref{fig:honeycomb}). Each carbon atom has four valence electrons. Three of the four valence electrons form covalent bonds with nearest-neighbor carbon atoms. The remaining valence electron is a $2p_z$ electron. This electron can freely move through the two-dimensional material. The elements silicon, germanium and tin also have two electrons in their outermost $s$ and $p$ shells and therefore these materials are potential candidates to also exhibit a graphene-like structure. Unfortunately, silicene, germanene, and stanene i.e. the silicon, germanium, and tin analogues of graphene do not occur in Nature and therefore these materials have to be synthesized or grown on carefully selected substrates \cite{Vogt2012, Fleurence2012, Li2014, Davila2014,Bampoulis2014}. 

The group IV 2D Xenes share many properties with graphene, but there are also a few remarkable differences \cite{Acun2015}. The most eye-catching difference is the structure; the graphene honeycomb lattice is fully planar, whereas the group IV 2D Xenes lattices are buckled, i.e. the two triangular sub-lattices are slightly displaced with respect to each other in a direction normal to the two-dimensional material \cite{Takeda1994,Guzman2007,Cahangirov2009}. As we will discuss later this broken sub-lattice symmetry allow the opening of a band gap if a transversal electric field is applied. Another difference is the spin-orbit coupling. The spin-orbit coupling, which scales as the fourth power of the atomic number, results in a tiny spin-orbit gap of only a few µeV in graphene, but in substantial spin-orbit gaps in silicene, germanene, and stanene  of $\sim$1.5 meV, $\sim$24 meV and $\sim$100 meV, respectively  \cite{Liu2011,Yao2007}. Owing to this spin-orbit gap, graphene as well as the other group IV 2D Xenes are strictly speaking two-dimensional topological insulators, which, as we will see later, are expected to exhibit the quantum spin Hall effect \cite{Kane2005,Kane2005_2}. In the case of graphene, however, the spin-orbit bandgap is so small that the quantum spin Hall effect is only expected to show up at extremely low temperatures.\\

\begin{figure}[H]
	\centering

	\includegraphics[trim=0cm 7.5cm 18cm 4cm,clip=true,width=0.8\textwidth]{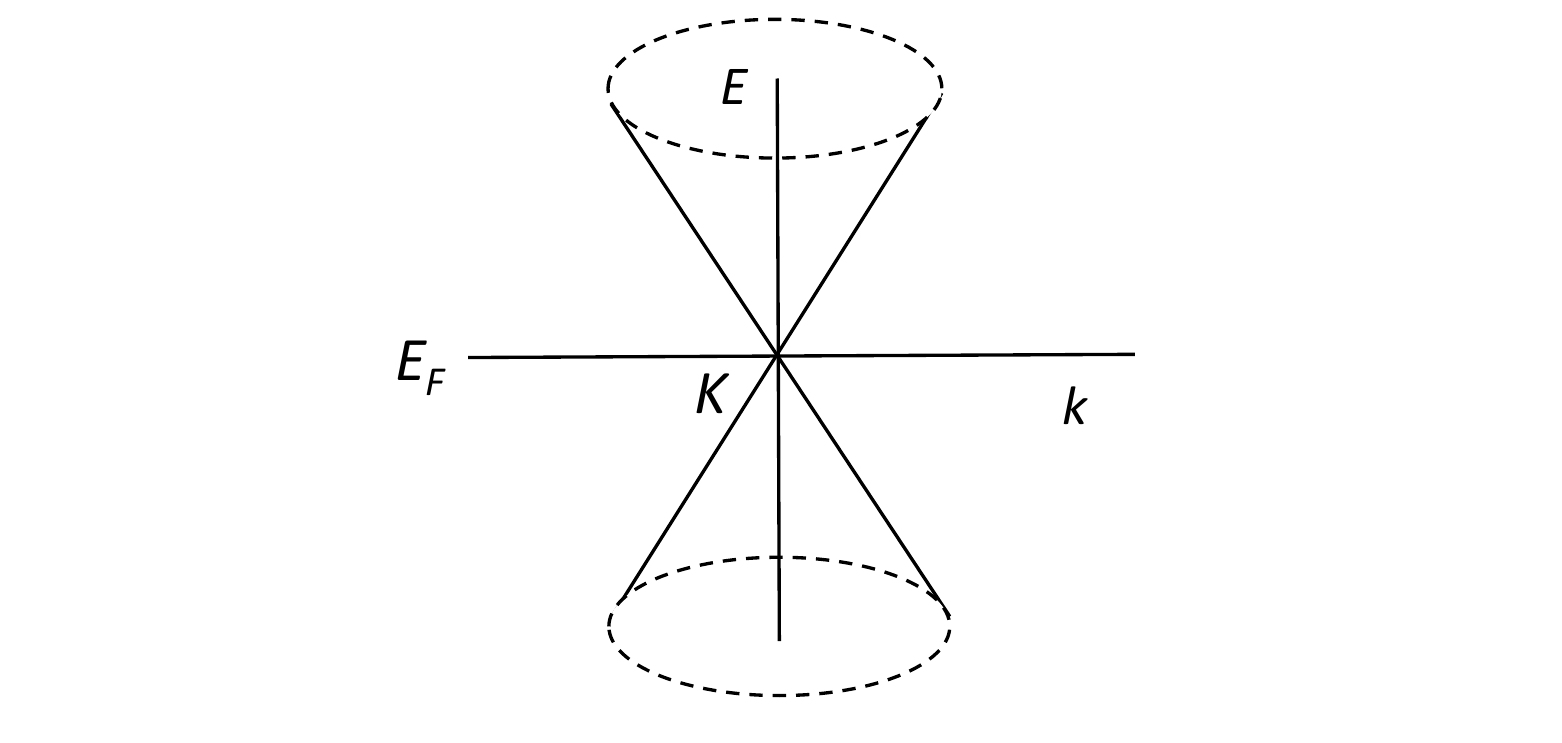}
	\caption{The electronic band structure of graphene at the $K$ valley.}
		\label{fig:diraccone}
\end{figure}

Despite the above-discussed differences between graphene and the other group IV 2D Xenes, simple tight binding calculations have revealed that the low-energy bands of all these two-dimensional materials are linear, i.e.,
\begin{equation}
	E = \pm \hbar v_{\text{F}} k
\end{equation}

where $\hbar$ is the reduced Planck constant, $v_F$ the Fermi velocity, $k$ the momentum of the electron with respect to the $K$ (or $K’$) point in the Brillouin zone and the sign $\pm$ refers to electrons or holes, respectively. Please note that the spin-orbit coupling has not been taken into account in these tight binding calculations. The linear dispersion relation, which gives rise to a Dirac cone in reciprocal space (see Figure \ref{fig:diraccone}), implies that the electrons in these materials behave as massless relativistic particles that obey the Dirac equation, i.e. the relativistic variant of the Schrödinger equation. The Fermi velocity is given by $v_\text{F} = \frac{1}{2}\sqrt{3} \frac{a}{\hbar}t$, where $a$ is the lattice constant and $t$ the in-plane hopping energy. For graphene, which has a lattice constant of 2.46 Å and an in-plane hopping energy of 3 eV, this results in a Fermi velocity of 1$\cdot 10^6$ m/s. The Fermi velocities of silicene, germanene and stanene are predicted to be comparable to the Fermi velocity of graphene.

The 2D Dirac equation is given by \cite{Neto2009}:
\begin{equation}
	i\hbar \frac{\partial \psi}{\partial t} = \left[ -i\hbar \left(\sigma_x 	\frac{\partial}{\partial x} + \sigma_y \frac{\partial}{\partial y}\right) + \sigma_z m	\right] \psi
\end{equation}
where $\sigma_x = \begin{pmatrix} 0 & 1 \\ 1 & 0 \end{pmatrix} $, $\sigma_y = \begin{pmatrix} 0 & -i \\ i & 0 \end{pmatrix}$, $ \sigma_z = \begin{pmatrix} 1 & 0 \\ 0 & -1 \end{pmatrix}$, $m$ is the mass of the quantum mechanical patricle and $\psi = \begin{pmatrix}
	\psi_A \\ \psi_B
\end{pmatrix}$ in which $\psi_{A,B}$ refer to the Bloch wave functions of the two triangular sub-lattices. 
The time independent Dirac equation for the massless low-energy electrons in graphene is given by,

\begin{equation}
	\begin{pmatrix}
		0 & -i\hbar v_F (\partial_x -i\partial_y) \\ 
		-i\hbar v_F (\partial_x + i\partial_y) & 0
	\end{pmatrix} 
	\begin{pmatrix}
		\psi_A \\ \psi_B
	\end{pmatrix}
	=
	E  \begin{pmatrix}
		\psi_A \\ \psi_B
	\end{pmatrix}
\end{equation}

Due to the symmetry of the honeycomb lattice, both sub-lattices contribute equally to the total wave function, only differing by a phase factor. This eigenvalue problem can be solved and provides the eigenvalues, $E_\pm =\pm \hbar v_\text{F} |k|$. For undoped graphene the Dirac point is located at the Fermi level.
The eigenfunctions that correspond to the eigenvalues $E_\pm =\pm \hbar v_F |k|$ are given by,
\begin{equation}
	\label{eq:eigenvalues}
	\psi = \begin{pmatrix}
		\psi_A \\ \psi_B
	\end{pmatrix}
	= 
	\frac{1}{\sqrt{2}}
	\begin{pmatrix}
		1 \\ \pm e^{i\theta}
	\end{pmatrix}
	e^{ikr}
\end{equation}

where $r$ is the position vector. The + sign refers to the conduction band and the - sign refers to the valence band, respectively. The angle between the crystal momentum, $p=\hbar k$ , and the $x$-axis is denoted by the aforementioned phase factor $\theta$. The two component vector term in Eq.\ref*{eq:eigenvalues} represents the pseudospin. This pseudospin is not a real spin, but can be considered as the vector projection of the real spin on the momentum vector of the electron \cite{Neto2009}. This pseudospin is locked to the momentum of the electron and is aligned parallel to the momentum for one valley (K) and antiparallel for the other valley ($K’$). This gives rise to another interesting property of graphene and the group IV 2D Xenes, which is referred to as the chirality of an electron. This chiral character of the electrons will be touched upon later.

\newpage
\subsection{The anomalous quantum Hall effect in graphene and group IV 2D Xenes}

\begin{figure}[H]
	\centering
	\includegraphics[trim=0cm 7cm 8cm 2cm,clip=true,width=\textwidth]{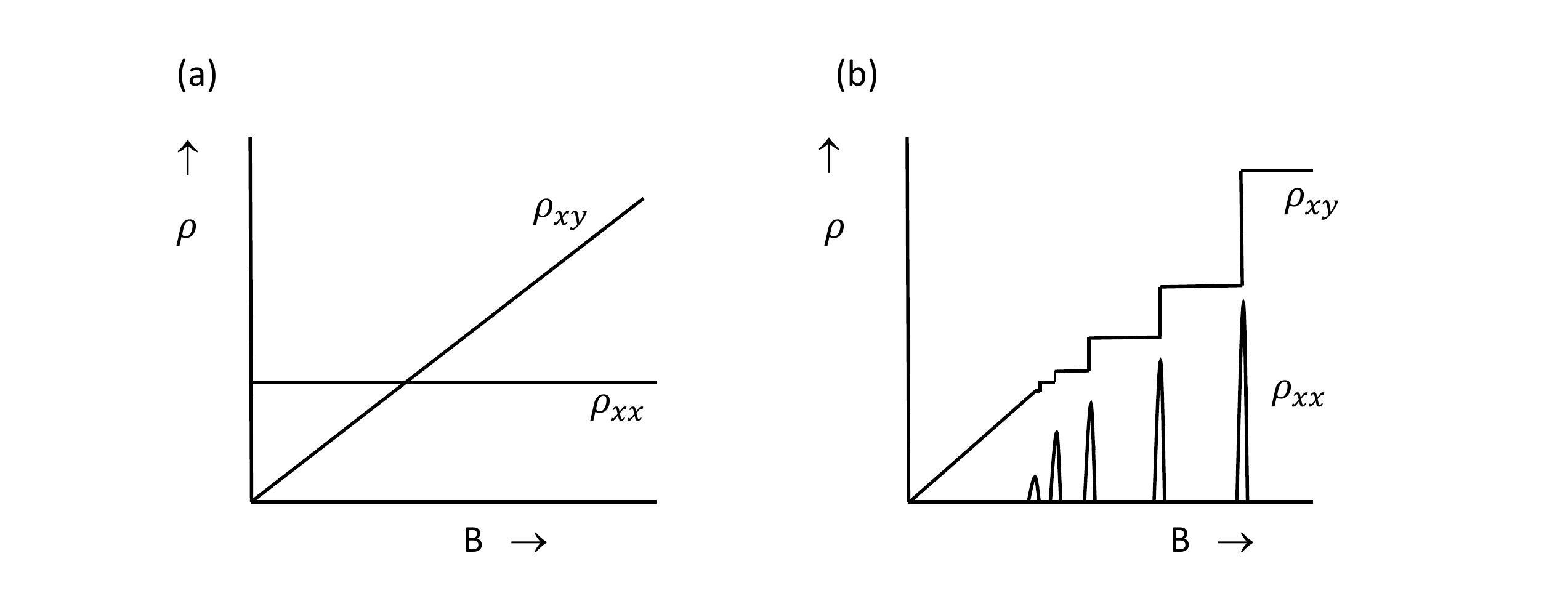}
	\caption{Plot of the longitudinal $\rho_{xx}$ and transversal, or Hall, resistivity ($\rho_{xy}$) versus magnetic field for the conventional Hall effect (a) and the integer quantum Hall effect (b).}
		\label{fig:hall}
\end{figure}

If an electric current flows through an electrical conductor and a magnetic field is applied in a direction perpendicular to this current, a voltage difference perpendicular to the applied electric current and applied magnetic field is produced. This effect, which stems from the Lorentz force that is experienced by the charge carriers, was discovered by Edwin Hall in 1879 and is referred to as the Hall effect. The Hall resistance provides information on the charge carrier density as well as the sign of the charge of the charge carriers. About a century later Klaus von Klitzing observed in two-dimensional electron gases at high magnetic fields and low temperatures that the Hall conductivity exhibits well-defined quantized steps. The applied external magnetic field causes the electrons to perform a circular motion with well-defined radii due to the wave nature of the electrons. In the absence of a magnetic field the density of states of a two-dimensional electron gas is a constant, i.e. independent of the energy of the electron. The application of a transverse magnetic field, however, results in a density of states that consists of a set of equidistant peaks, the so-called Landau levels \cite{Klitzing1986}. The Landau levels $E_{\text{LL}}$ are given by,
\begin{equation}
	E_{\text{LL}} = \hbar \omega_\text{c} \left( n + \frac{1}{2}\right)
\end{equation}
where $n=0,1,2,...$ is the quantum number, $\omega_{\text{c}}=eB/m$ is the cyclotron frequency, $e$ the elementary charge and $B$ the applied transverse magnetic field. At the edges of the sample the electrons, which all orbit in the same clockwise (or anti-clockwise) direction, bounce back and perform a skipping trajectory leading to a net flow of current along the edges of the sample. The current that flows along the edges of the sample is robust and dissipation-less because back-scattering is forbidden. The transversal conductivity, $\sigma_{xy}$ , is quantized in integer units of $e^2/h$. The inverese quantity $h/e^2$ is referred to as the quantum of resistance or von Klitzing constant and is nowadays used as the standard for measuring resistivity, as the constant can be measured with a precision better than a few parts in ten billion [18]. Interestingly, the presence of disorder, i.e. defects or impurities, has no effect on this quantization. In fact, in the disorder-free case the quantized steps in the Hall conductivity are expected to disappear completely. The longitudinal resistivity, $\rho_{xx}$, is zero as long as the transversal conductivity resides on a plateau level. If the transversal conductivity jumps to the next plateau a spike in the longitudinal resistivity occurs (see Figure \ref{fig:hall}).

The electrons in two-dimensional Dirac materials behave differently than in a conventional two-dimensional electron gas. First, the energy dispersion relation of the low energy electrons in a 2D Dirac material is not quadratic in the momentum, but linear. Second, the electrons in a 2D Dirac material are chiral. The chiral character of the electrons makes that the pseudospin is locked to the momentum. The unit cell consists of two atoms, one atom belongs to the triangular sub-lattice A, whereas the other atom belongs to the triangular sub-lattice B. The wave function of the electrons in a 2D Dirac material is composed of two Bloch waves, which are related to the two triangular sub-lattices. For symmetry reasons the relative contribution of the Bloch waves of both sub-lattices should be equal, which means that the wave functions of the two Bloch waves only differ by a phase factor. This phase factor, which has been introduced in the  previous section, is referred to as the pseudospin and is locked to the momentum of the electron. The linear dispersion of the electrons and their chiral character result in Laudau levels that are proportional to the square root of the magnetic field and the quantum number $n$\cite{Novoselov2006,Katsnelson2007},
\begin{equation}
	\label{eq:landau}
	E_{\text{LL}} =  \pm \sqrt{ 2 |e| B \hbar v_\text{F}^2 \left( n + \frac{1}{2} \pm \frac{1}{2} \right)}
\end{equation}
where $n=0,1,2,...$ . The $\pm$ term is due to the chiral character of the electrons. A two-dimensional material that hosts Dirac fermions exhibits a zero-energy state (the case of $n=0$  and the minus sign in equation \ref{eq:landau}). The existence of a zero-energy Landau level results in an anomalous quantum Hall effect with half-integer quantization of the Hall conductivity rather than an integer quantization. The Hall conductivity for graphene is given by\cite{Novoselov2006,Katsnelson2007},

\begin{equation}
	\sigma_{xy} = \pm \frac{4e^2}{h} \left( n + \frac{1}{2} \right)
\end{equation}

where the factor of 4 arises because of the spin and valley degeneracies. The anomalous quantum Hall effect is one of the most prominent signatures of a two-dimensional Dirac material. The anomalous quantum Hall effect has been convincingly demonstrated for graphene, but not yet for silicene, germanene and stanene. 

\subsection{Spin-orbit coupling and the quantum spin Hall effect}

So far, we have assumed that graphene and the other group IV 2D Xenes are gapless, which is strictly speaking not correct. Owing to spin-orbit coupling these 2D materials possess a spin-orbit gap that varies from a few $\upmu$eV for graphene to $\sim$100 meV for stanene. In order to observe the tiny spin-orbit gap in graphene, experiments at extremely low temperatures are required and therefore the spin-orbit gap in graphene can often safely be ignored. The latter is, of course, not true for the other group IV 2D Xenes. In order to elaborate on the spin-orbit coupling in these materials we consider the following Dirac Hamiltonian, $ H_{\xi, s}$,

\begin{equation}
	\label{eq:so}
	\begin{pmatrix}
		\xi s\lambda_{\text{so}} & -i\hbar v_F (\xi\partial_x -i\partial_y) \\ 
		-i\hbar v_F (\xi\partial_x + i\partial_y) & -\xi s\lambda_{\text{so}}
	\end{pmatrix} 
	\begin{pmatrix}
		\psi_A \\ \psi_B
	\end{pmatrix}
	=
	E  \begin{pmatrix}
		\psi_A \\ \psi_B
	\end{pmatrix}
\end{equation}

where $\lambda_{\text{so}}$ is the spin-orbit gap, $\xi(\pm 1) $ is the valley index and $s (\pm 1)$ is the spin index \cite{Motohiko2015}. The dispersion relation of the energy bands near the $K$ and $K'$ points of the Brillouin zone is given by,

\begin{equation}
	E(k) = \pm \sqrt{(\hbar v_F k)^2 + (\xi s \lambda_{\text{so}})^2}
\end{equation}

The effective mass at the $K$ and $K’$ points of the Brillouin zone, also referred to at the Dirac mass can be determined using,  
\begin{equation}
	m_{\text{eff}} = \frac{\pm \xi s \lambda_{\text{so}}}{v_F^2}
\end{equation}

In order to obtain information on the topological properties of 2D materials, a thorough understanding of the electronic band structure is required. We will first determine the Berry connection and curvature, as these quantities are needed to determine the Chern number. Let us first define the Berry connection, which can be considered as a gauge potential. For a two-dimensional system the Berry connection for any insulating state is defined as,
\begin{align}
	A_x (\mathbf{k}) = -i\braket{\psi (\mathbf{k}) |\partial_x | \psi (\mathbf{k})  } \\
	A_y (\mathbf{k}) = -i\braket{\psi (\mathbf{k}) |\partial_y | \psi (\mathbf{k})  }
\end{align}
This Berry connection is related the gauge field, also referred to as the Berry curvature, $\Omega(k)$, via,
\begin{subequations}
	\begin{align}
		\Omega (\mathbf{k}) &= \nabla \times A(\mathbf{k}) \\
		\Omega (\mathbf{k}) &= \frac{\partial}{\partial k_x} (-i\braket{\psi (\mathbf{k}) |\partial_y | \psi (\mathbf{k})  }) - \frac{\partial}{\partial k_y} (-i\braket{\psi (\mathbf{k}) |\partial_x | \psi (\mathbf{k})  } )
	\end{align}	
\end{subequations}

The Chern number is defined as the integral of the Berry curvature over the first Brillouin zone, i.e.,
\begin{equation}
	C = \frac{1}{2\pi} \int \Omega (\mathbf{k})d\mathbf{k}
\end{equation}
This Chern number is an integer and can be considered as the total flux of the gauge field. The Chern number determines, amongst others, the contribution of an electronic band to the Hall conductivity $\sigma_{xy} =Ce^2/h$, when the band is completely filled. This is an interesting observation as the properties of the edge of a topological insulator are in principle governed by the properties of the material in the bulk. The latter is referred to as the bulk-boundary principle. For graphene and the other group IV 2D Xenes the Berry curvature is strongly peaked at the valley maxima and minima. For the Dirac Hamiltonian given in Eq. \ref{eq:so} the Berry curvature is given by,
\begin{equation}
	\label{eq:berrycurvature}
	\Omega_{\xi, s}(k) = -\xi \frac{\xi s \lambda_{\text{so}}}{2\left[(\hbar v_\text{F} k)^2+ (\xi s \lambda_{\text{so}})^2\right]^{3/2}}
\end{equation}
Using Eq. \ref{eq:berrycurvature}, we find for the valley Chern number,
\begin{equation}
	C_{\xi, s} = -\frac{\xi}{2}\frac{\xi s \lambda_{\text{so}}}{|\xi s \lambda_{\text{so}}|}
\end{equation}

In the absence fo an external electric field, we find the following valley Chern numbers:

\begin{equation}
	C_{\xi = +1 \text{,} s = + 1} = -\frac{1}{2} \text{ , }
	C_{\xi = +1 \text{,} s = - 1} = +\frac{1}{2} \text{ , }
	C_{\xi = -1 \text{,} s = + 1} = -\frac{1}{2} \text{ , }
	C_{\xi = -1 \text{,} s = - 1} = +\frac{1}{2} 
\end{equation}

\begin{figure}[H]
	\centering
	\includegraphics[trim=3cm 7cm 9cm 4cm,clip=true,width=\textwidth]{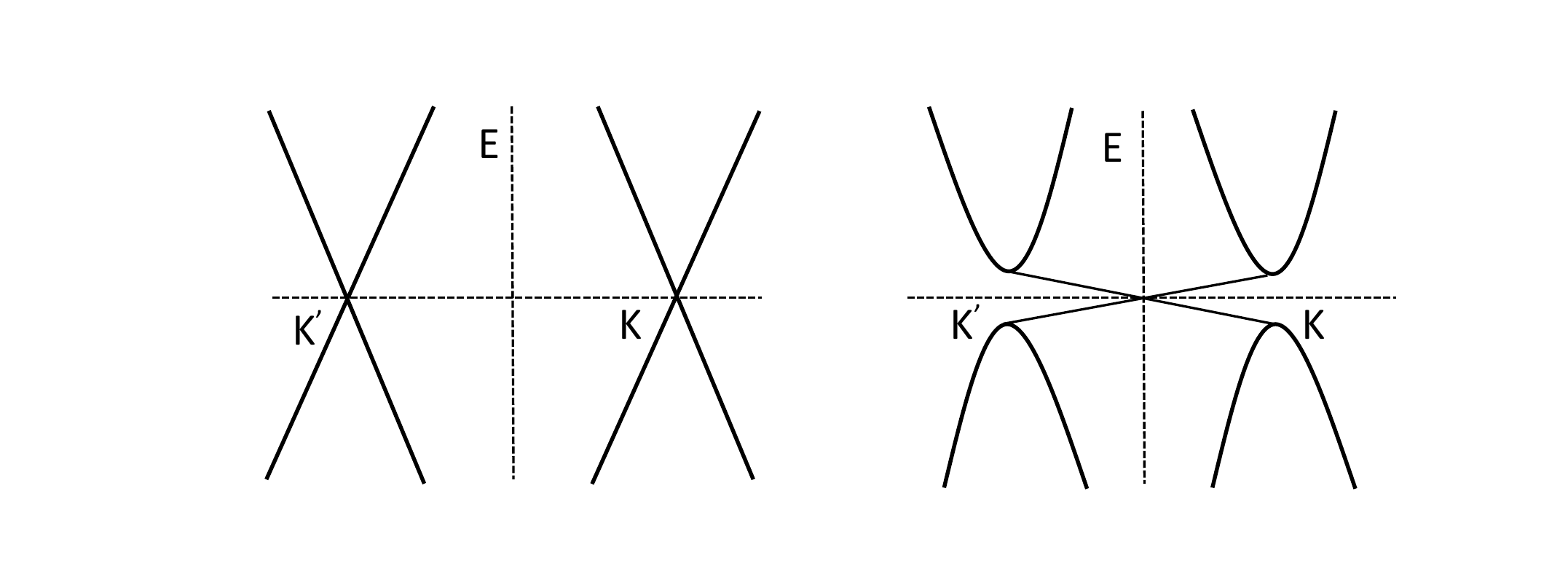}
	\caption{(a) Electronic band structure of graphene without spin-orbit coupling near the Fermi level. (b) Electronic band structure of graphene with spin-orbit coupling near the Fermi level.}
	\label{fig:spinorbit}
\end{figure}

For the total Chern number, we find the following value:
\begin{equation} 
	C_\text{tot} = C_{\xi = +1 \text{,} s = + 1} + C_{\xi = +1 \text{,} s = - 1} + C_{\xi = -1 \text{,} s = + 1} + C_{\xi = -1 \text{,} s = - 1} = 0
\end{equation} 
On the other hand, the total spin Chern number, is unequal to zero:
\begin{equation}
	C_\text{spin} = \left(C_{\xi = +1 \text{,} s = + 1} + C_{\xi = -1 \text{,} s = + 1}\right) - \left( C_{\xi = +1 \text{,} s = - 1} + C_{\xi = -1 \text{,} s = - 1} \right) = -2.
\end{equation} 
These Chern numbers are associated with a quantum spin Hall insulator. The spin-orbit coupling results in an internal magnetic field that couples to the spin of the electrons. Spin-up electrons are pushed to one side of the sample, whereas spin-down electrons are pushed to the opposite direction. This asymmetry give rise to two spin-polarized conduction channels at the edges of the 2D material that propagate in opposite directions, the so-called gapless and topologically protected helical edge modes, as shown in figure~\ref{fig:spinorbit}. The quantum spin Hall effect is characterized by a vanishing charge Hall conductance ($C_{\text{tot}} = 0$) and a quantized spin Hall conductance ($C_{\text{spin}}=-2$) of $2e/4\pi$ (an electron with charge $e$ carries a spin $\hbar/2$ and therefore the spin Hall conductance becomes  $\frac{2e^2}{h}\frac{\hbar}{2e} =e/2\pi$). 

In 2005 Kane and Mele \cite{Kane2005,Kane2005_2} were the first to point out that graphene is a $Z_2$ topological insulator and should exhibit the quantum spin Hall effect. In Figure 4 the electronic band structure of graphene without and with spin-orbit coupling is shown. This spin-orbit results in a two counter-propagating spin-polarized topologically protected edge states. Silicene, germanene, and stanene, which also fall in the class of $Z_2$  topological insulators, have a much larger spin-orbit gap and therefore the quantum spin Hall effect should be experimentally accessible. Unfortunately, there is no experimental evidence yet for the existence of the quantum spin Hall effect in the group IV 2D Dirac materials.\\

In 2006 Bernevig, Hughes and Zhang predicted that the quantum spin Hall effect can also occur in CdTe/HgTe/CdTe quantum wells \cite{Bernevig2006}. By varying the thickness of the quantum well, the band structure can be switched from a normal to an “inverted” type at a critical thickness. Shortly after this prediction the QSHE state in a CdTe/HgTe/CdTe quantum well was experimentally confirmed by König \textit{et al}. \cite{Konig2007}. These authors demonstrated at low-temperatures ($T< 1.4$ K) the presence of  an edge conductance   that only exists beyond the critical layer thickness of the quantum well. In 2017 Reis \textit{et al}. \cite{Reis2017} pointed out that bismuthene synthesized on a SiC substrate is an appealing candidate to exhibit the quantum spin Hall effect. SiC(0001) is an insulator that exhibits a ($\surd3\times\surd$3)R30\textdegree{} structure with a lattice constant of 5.35 Å, which is substantially larger than the lattice constant of buckled Bi(111) bilayers. Owing to this difference in lattice constant the honeycomb lattice of bismuthene becomes fully planar. This planarity is advantageous as an external electric field will not result in a charge shift from one triangular sub-lattice to other sub-lattice of the honeycomb lattice of bismuthene. In addition, if the bismuthene layer would have been buckled it becomes very susceptible to substrate interactions. Even a relatively weak interaction with the substrate can change the buckling and alter the electronic bandgap structure \cite{Zhang2016}. Reis \textit{et al}. \cite{Reis2017} pointed out that bismuthene is different from graphene in the sense that in graphene the low energy electrons are the $\pi$ electrons, whereas in bismuthene the low-energy electrons are the $p_x$- and $p_y$-orbitals. Scanning tunneling spectroscopy measurements revealed that the edges of bismuthene on SiC(0001) are metallic, while the interior of the bismuthene has a bandgap of about 0.8 eV. In order to prove the topological nature of the edges states of bismuthene electronic transport measurements are, however, still required.\cite{Stuhler2020}

\subsection{Application of a transversal electric field and topological phase transitions}

\begin{figure}[H]
	\centering
	\includegraphics[trim=0cm 0cm 0cm 0cm,clip=true,width=\textwidth]{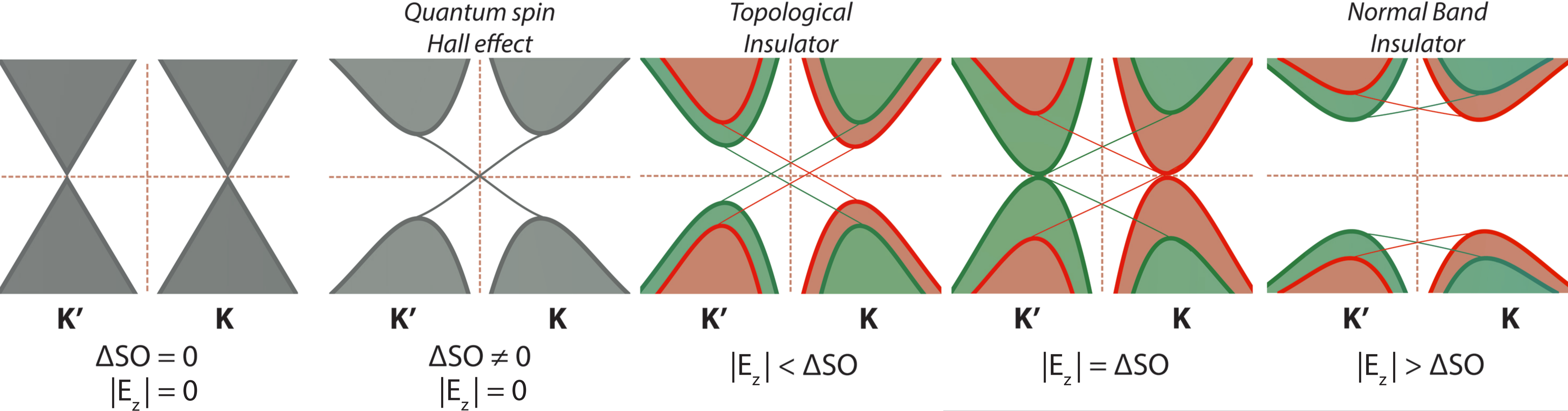}
	\caption{Schematic diagram of the electronic band structure of a 2D Dirac material with a buckled honeycomb lattice near the $K$ and $K’$ points of the Brillouin zone. From left to right: electronic band structure without a spin-orbit gap, with a spin-orbit gap, applied electric field smaller than the critical value, applied electric field equal to the critical value and applied electric field larger than the critical value. Red and green refer to spin up and spin down bands, respectively. (Reprinted with permission from Institute of Physics, A. Acun \textit{et al}., J. Phys. Cond. Matt. \textbf{27}, 443002 (2015)).}
		\label{fig:QSHE}
\end{figure}

In case a transversal electric field is applied to graphene, the electronic band structure is not affected, but only the Fermi level will be shifted. The gating results in an induced charge per unit area that is given by $C_\text{g} V_\text{g}$, where $V_\text{g}$ is the gate voltage and $C_\text{g}$ is the capacitance per unit area. If we assume that the pristine graphene is undoped, we can determine the exact shift of the Fermi level due to the gating by integrating the density of states over the relevant energy window,
\begin{equation}
	n = C_gV_g = \int_{0}^{E_\text{F}}\frac{4|E|}{\pi(\hbar v_\text{F})^2}dE
\end{equation}
As discussed above, there is another salient difference between graphene and the other group IV 2D Xenes: the honeycomb lattice of graphene is perfectly planar, but the honeycomb lattices of the other group IV 2D Xenes are buckled. This broken symmetry allows to transfer charge from one triangular sub-lattice to the other sub-lattice via, for instance, the application of a transversal electric field. If a transversal electric field, $E_\text{z}$, is applied the bandgap in a group IV 2D Xene is given by \cite{Motohiko2015},
\begin{equation}
	\label{eq:bandgap}
	\Delta_{\xi,s} = \xi s \lambda_{\text{so}} - \frac{\delta}{2} E_\text{z}
\end{equation}
where $\updelta$ is the buckling, i.e. the vertical displacement of the two triangular sub-lattices. It is immediately clear from Eq. \ref{eq:bandgap} that the external electric field results in an asymmetry of the band gap for the different valleys and spins (see Figure \ref{fig:QSHE}). With increasing electric field the bandgap first decreases until it completely closes at a critical field $E_\text{c} = \pm (2\lambda_{\text{so}}/\delta)$. For electric fields exceeding this critical field the bandgap reopens again. For electric fields smaller than the critical field the 2D material is a quantum spin Hall insulator, but for electrical fields exceeding the critical field the material is a normal band insulator. The valley Chern numbers can be calculated using,
\begin{equation}
	\label{eq:chern_so}
	C_{\xi,s} = -\frac{\xi}{2}\left( \frac{\Delta_{\xi,s}}{|\Delta_{\xi,s}|}\right) = -\frac{\xi}{2}\left( \frac{\xi s \lambda_{\text{so}}- \frac{\delta}{2}E_\text{z}}{|\xi s \lambda_{\text{so}}- \frac{\delta}{2}E_\text{z}|}\right)
\end{equation}
%
Using Eq. \ref{eq:chern_so}, we find for electric fields larger than the critical field, the following Chern numbers are found:
\begin{equation}
	C_{\xi = +1 \text{,} s = + 1} = -\frac{1}{2} \text{ , }
	C_{\xi = +1 \text{,} s = - 1} = -\frac{1}{2} \text{ , }
	C_{\xi = -1 \text{,} s = + 1} = +\frac{1}{2} \text{ , }
	C_{\xi = -1 \text{,} s = - 1} = +\frac{1}{2}
\end{equation}
In this case both the total Chern number as well as the total spin Chern number are 0 and therefore the material behaves now as a normal band insulator. The total valley Chern numbers are -1 and +1, respectively, and therefore the material can also be classified as quantum valley Hall insulator.

\section{Bilayer graphene and twisted bilayer graphene}

\begin{figure}[H]
	\centering

	\includegraphics[trim=0cm 0cm 0cm 0cm,clip=true,width=0.8\textwidth]{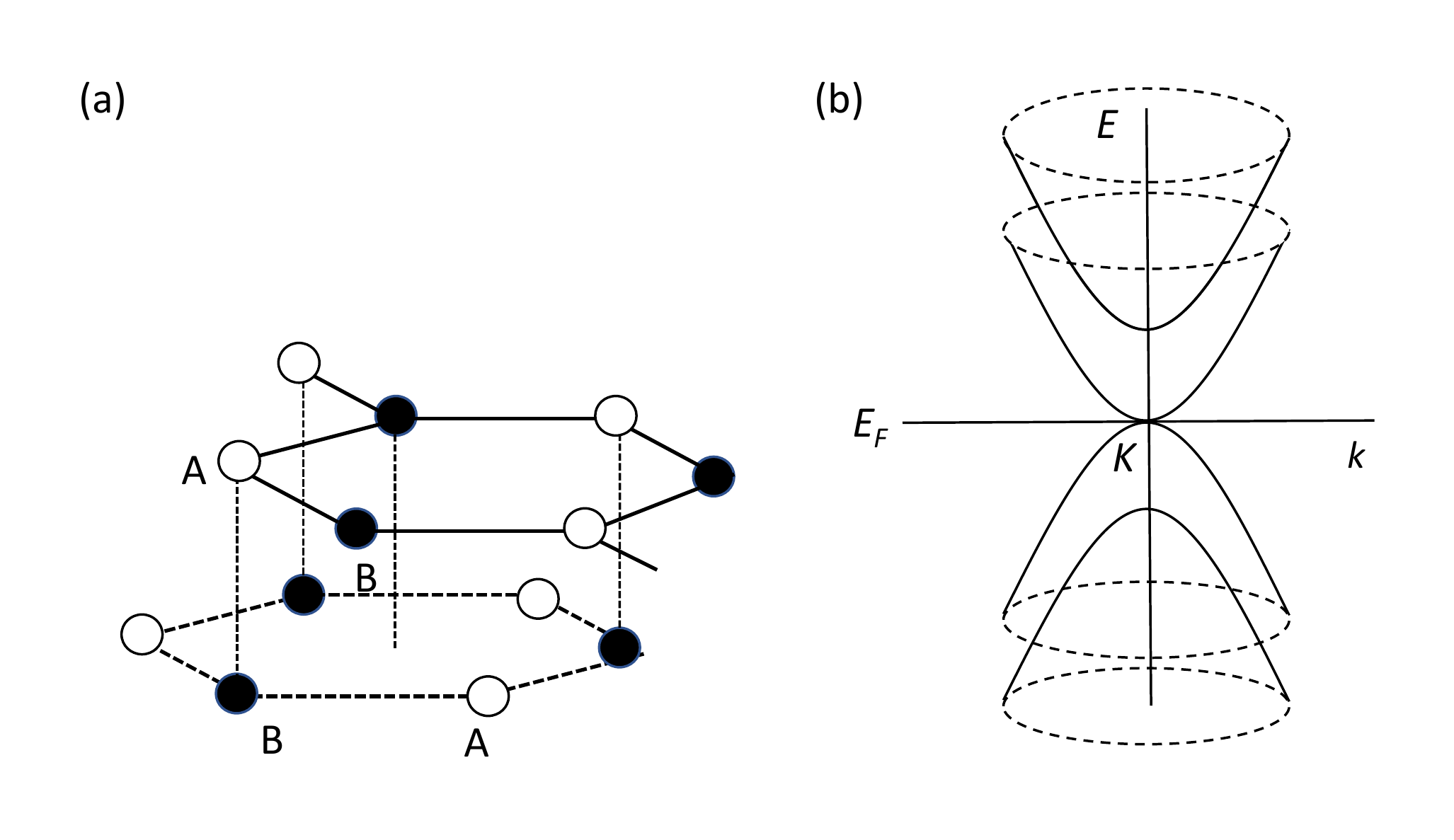}
	\caption{(a) Bernal (AB) stacked bilayer graphene. (b) Electronic band structure of the low-energy bands of Bernal stacked bilayer graphene near the Fermi level.}
		\label{fig:bilayergraphene}
\end{figure}

Next we will briefly elaborate on the electronic band structure of stacked bilayers of graphene. We will first consider the lowest energy configuration, which is the so-called AB stacked or Bernal stacked graphene. In AB stacked bilayer graphene the carbon atoms of sub-lattice A of the top layer sit on-top of the carbon atoms of sub-lattice B of the bottom layer. The carbon atoms of sub-lattice B of the top layer are located exactly above the hexagons of the bottom graphene layer (see Figure \ref{fig:bilayergraphene}a). The electronic band structure of AB stacked bilayer graphene is different from the electronic band structure of single layer graphene. Like graphene, AB stacked bilayer graphene is also a semimetal, but the low-energy electrons in the bilayer are not massless. The carbon atoms that are located on top of each other, which are often referred to as dimer sites, form bonding and anti-bonding orbitals. There are in principle four low-energy bands, which are denoted $E_{2}^{\pm}$, respectively (see Figure \ref{fig:bilayergraphene}b). The $E_{1,2}^{\pm}$ bands are gapped at the $K$ and $K’$ points of the surface Brillouin zone and will not be further discussed here. The $E_{1}^{\pm}$ bands, which touch at the Fermi level, can be approximated by \cite{McCann2013},
\begin{equation}
	E_1^{\pm} \approx \frac{1}{2}t_{\perp} \left[ \sqrt{1 + \frac{4v^ 2p^2}{t_{\perp}^2}} - 1\right]
\end{equation}
where $v$ is the velocity of the electron, $p$ the momentum of the electron and $t_{\perp}$ the interlayer hopping energy ($\approx 0.22$ eV for graphene \cite{McCann2013}). For small momenta the dispersion relation is quadratic, i.e. $E_1^{\pm} \approx \pm \frac{v^2p^2}{t_{\perp}}$, whereas for large momenta a linear dispersion relation is found, $E_1^{\pm} \approx \pm vp$. The crossover between the two regimes occurs at $p \approx \frac{t_{\perp}}{2v}$.

Subsequently we consider a twisted bilayer graphene, where the top layer makes a rotation angle, $\theta$ with respect to the bottom layer (see Figure 7). A twist angle $\theta = 0$\textdegree{} corresponds to the AB stacking, while an angle $\theta = 60$\textdegree{} corresponds to the AA stacking. A twist angle of exactly 30\textdegree{} is also special as it results in a two-dimensional quasi-crystal with 12-fold rotational symmetry, but no translation symmetry. The twist results in a moiré structure with a periodicity of $\lambda = a/2\sin(\theta/2) $, where $a$ is the lattice constant of graphene.

\begin{figure}[H]
	\centering
	\includegraphics[trim=0cm 0cm 0cm 0cm,clip=true,width=\textwidth]{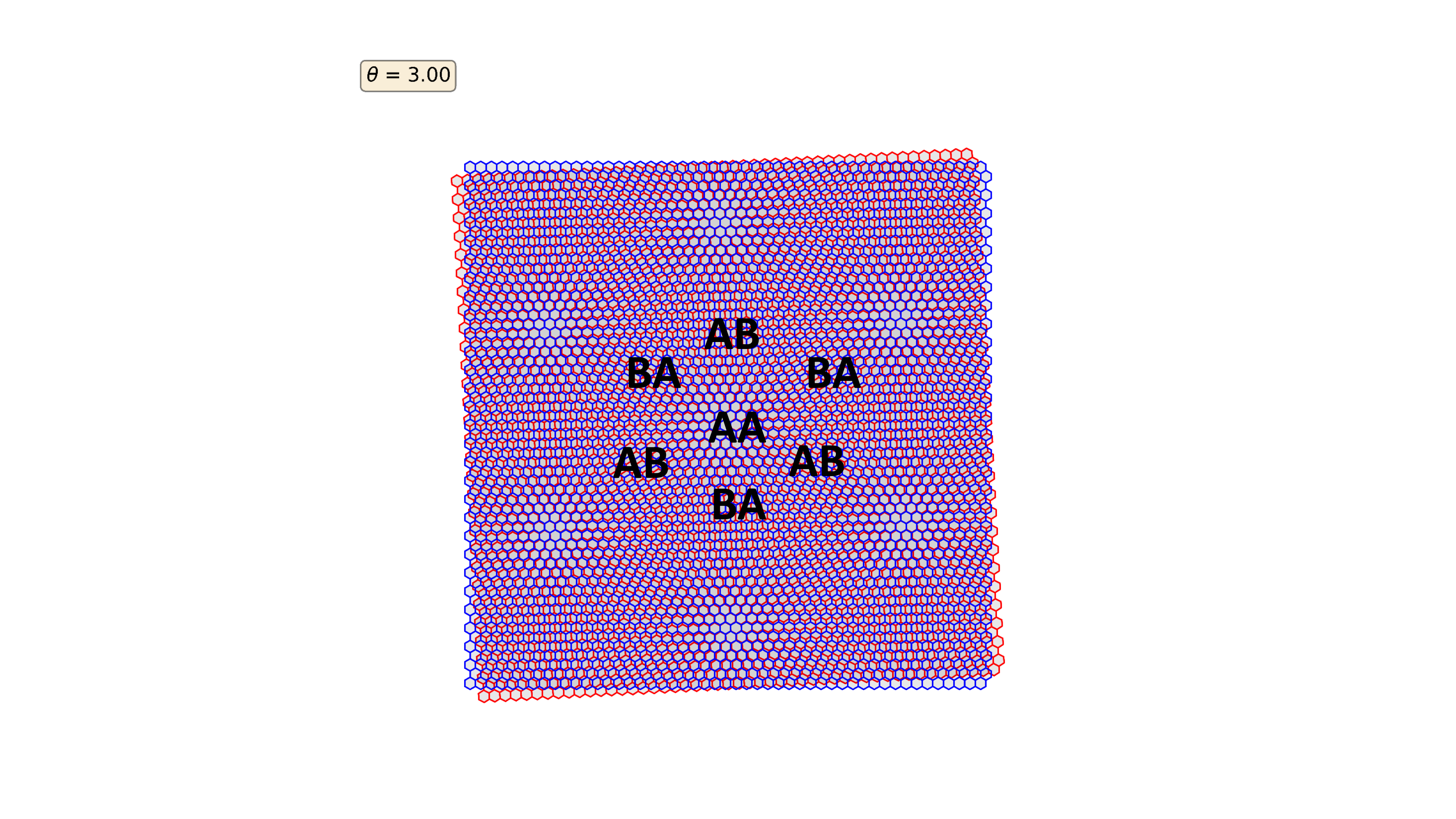}
	\caption{Twisted bilayer graphene. The twist angle is 3\textdegree{}. There three different stacking domains, AA, AB and BA.}
		\label{fig:twisted}
\end{figure}

\begin{figure}[H]
	\centering

	\includegraphics[trim=0cm 0cm 0cm 0cm,clip=true,width=0.7\textwidth]{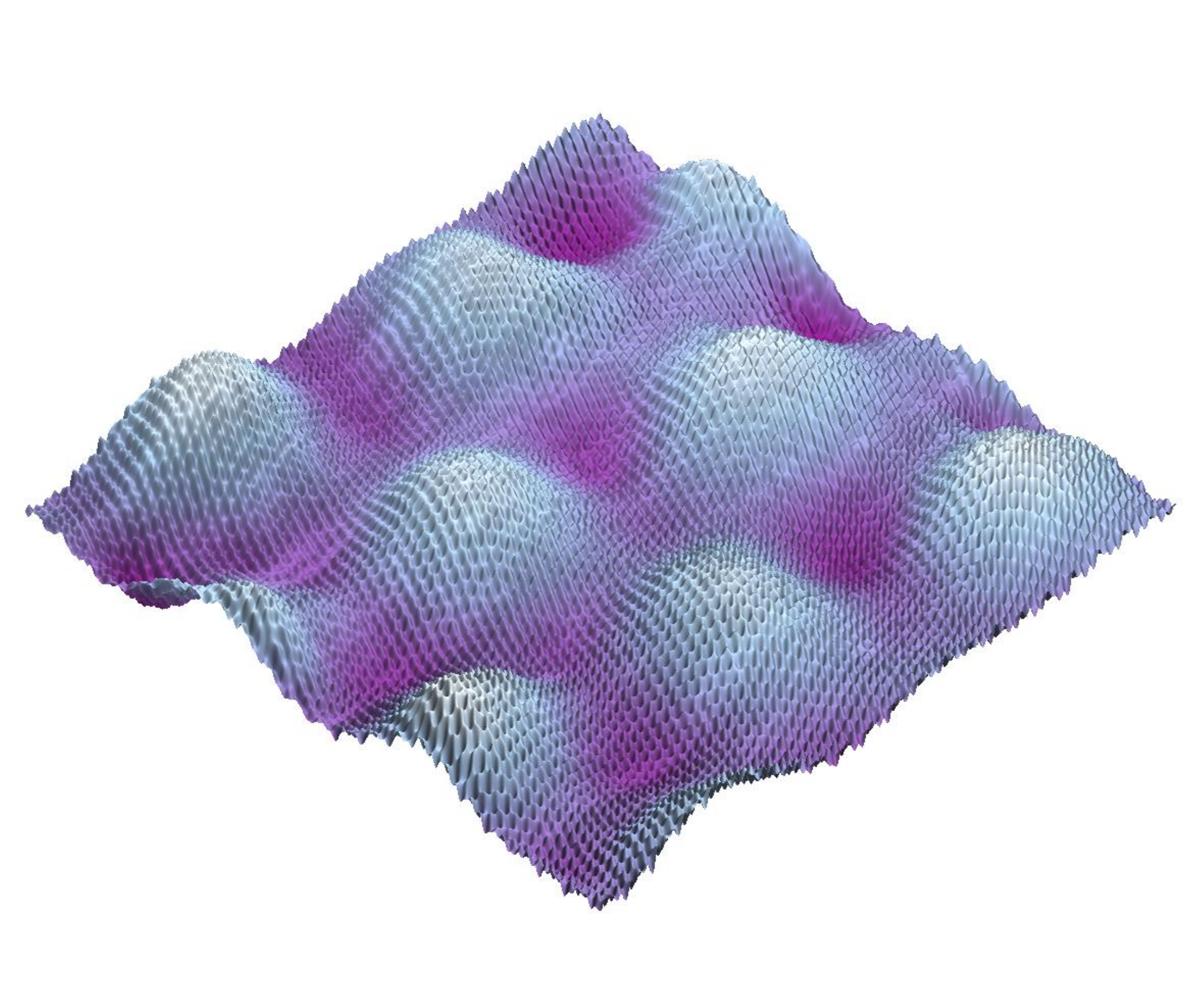}
	\caption{Scanning tunneling microscopy image of a twisted bilayer graphene. The twist angle is 2.30. The periodicity of the moiré structure is 6.5 nm. (reprinted with permission from American Physical Society, Q. Yao \textit{et al}., Phys. Rev. B \textbf{95}, 245116 (2017)).}
		\label{fig:twisted2}
\end{figure}

\begin{figure}[H]
	\centering

	\includegraphics[trim=6cm 4cm 6cm 4cm,clip=true,width=\textwidth]{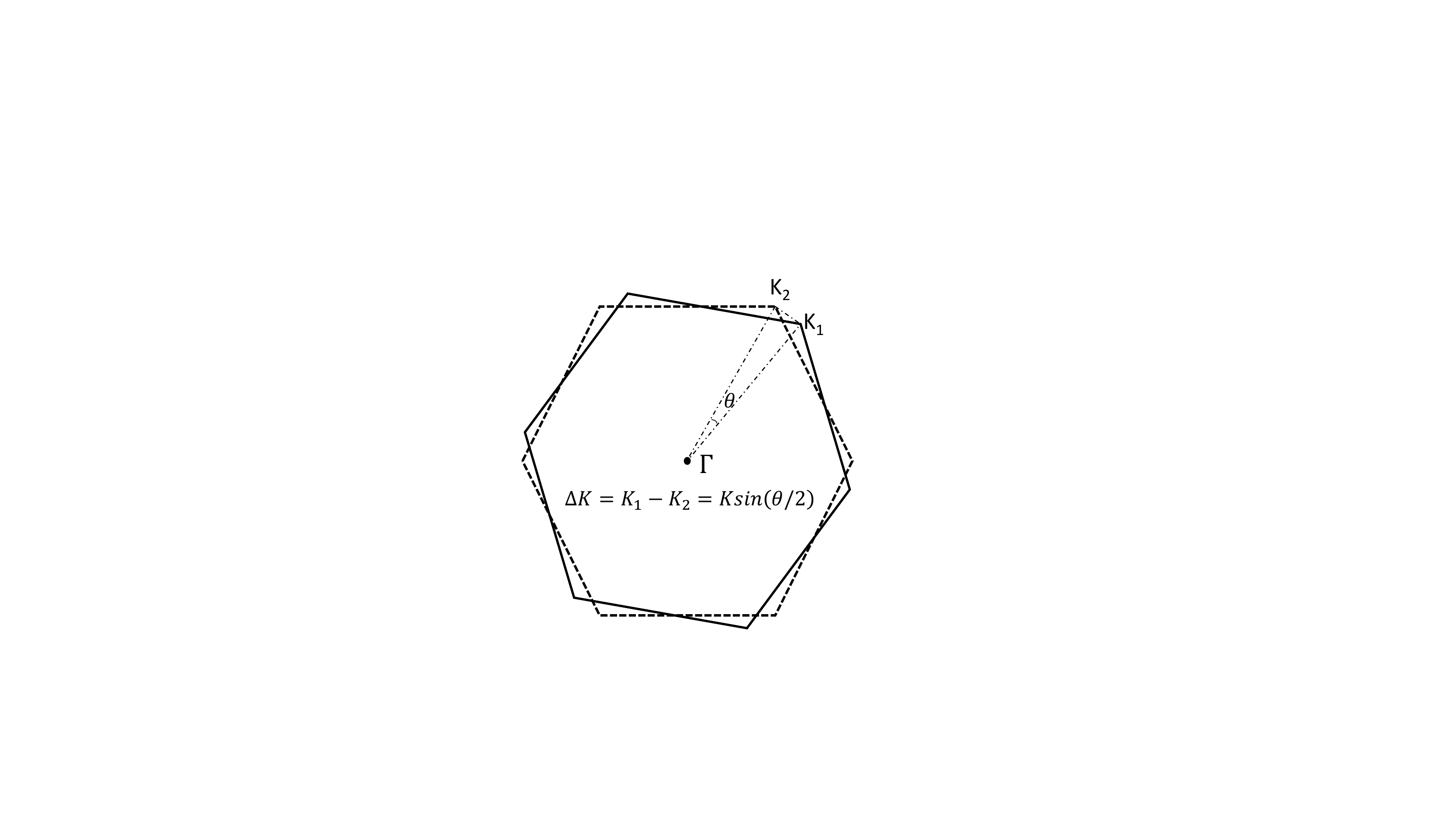}
	\caption{Reciprocal lattice of twisted bilayer graphene. The solid and dotted lines refer to the top and bottom graphene layers, respectively. The twist angle is denoted by $\theta$.}
		\label{fig:twist_k}
\end{figure}

\begin{figure}[H]
	\centering

	\includegraphics[trim=2cm 7cm 12cm 2cm,clip=true,width=\textwidth]{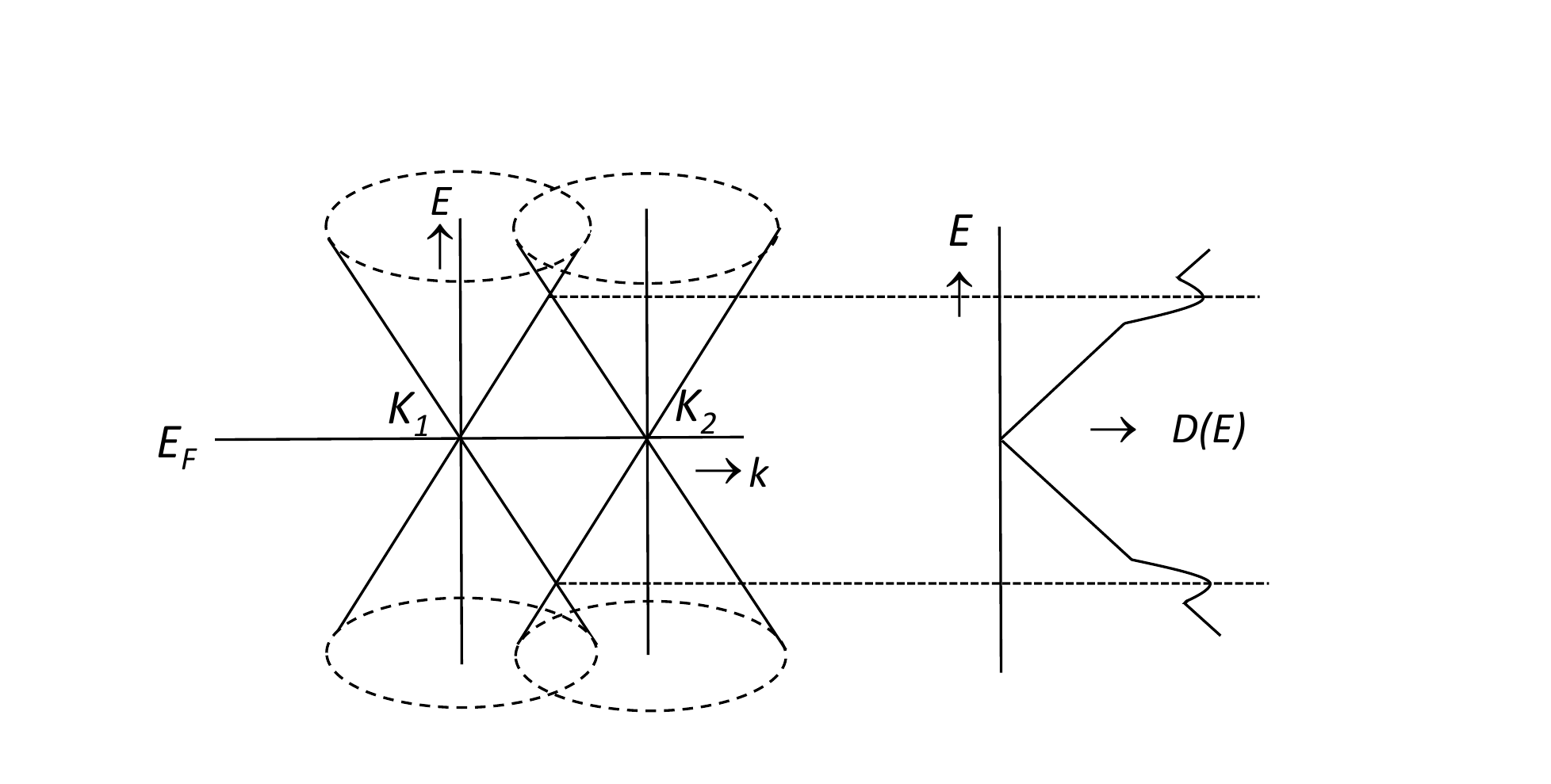}
	\caption{The crossing of two Dirac cones result in the development of two van Hove singularities. }
		\label{fig:vanHove}
\end{figure}

The hexagonal moiré structure consist of an ordered arrangement of AB, BA, AA and BB stacked regions (Figures \ref{fig:twisted} and \ref{fig:twisted2}).  One might naively assume that this periodic modulation results in the opening of a bandgap as for ‘normal’ electrons. In the case of massless Dirac fermions the chirality, however, prevents the opening of a bandgap. In the case of graphene on hexagonal boron nitride the periodic moiré potential results in the appearance of new Dirac points \cite{Yankowitz2012}. The twist of the top graphene layer leads to a shift of the two Dirac points in momentum space. The shift in momentum space of the Dirac points with respect to the $K$ point, $\Delta K$, is given by $\Delta K = \pm K\sin(\theta/2)$  (see Figure \ref{fig:twist_k}) \cite{Li2010,Hove1953,Ohta2006,Castro2007,Li2009,Trambly2010,Luican2011}. The crossing of the Dirac cones in the vicinity of the Fermi level leads to a ‘flat’ region in the energy dispersion relation, i.e. a Van Hove singularity (see Figure \ref{fig:vanHove}). 

A prerequisite for the formation of Van Hove singularities in bilayer graphene is the presence of a non-zero interlayer coupling. For a vanishing interlayer coupling the low energy electronic band structure of bilayer graphene, of course, reduces to that of a single layer of graphene. In the case of twisted bilayer graphene the Van Hove singularities can be brought arbitrarily close to the Fermi level. If the Fermi level lies within the flat bands, the Coulomb interaction is much larger than the kinetic energy of the electrons, which can drive the system into various strongly correlated electron phases \cite{Cao2018_1,Cao2018_2,Padhi2018}. 

\newpage
In Figure 11 two examples of twisted bilayer graphene are shown. The first example has a twist angle of 3.3\textdegree{} and two well-defined Van Hove singularities at -120 meV and 198 meV, respectively. The second example has a twist angle of 1.6\textdegree{} and two Van Hove singularities at -24 meV and 122 meV, respectively \cite{Yao2020}.

\begin{figure}[H] 
	\centering
	\vspace{1cm}
	\includegraphics[trim=0cm 0cm 0cm 0cm,clip=true,width=0.85\textwidth]{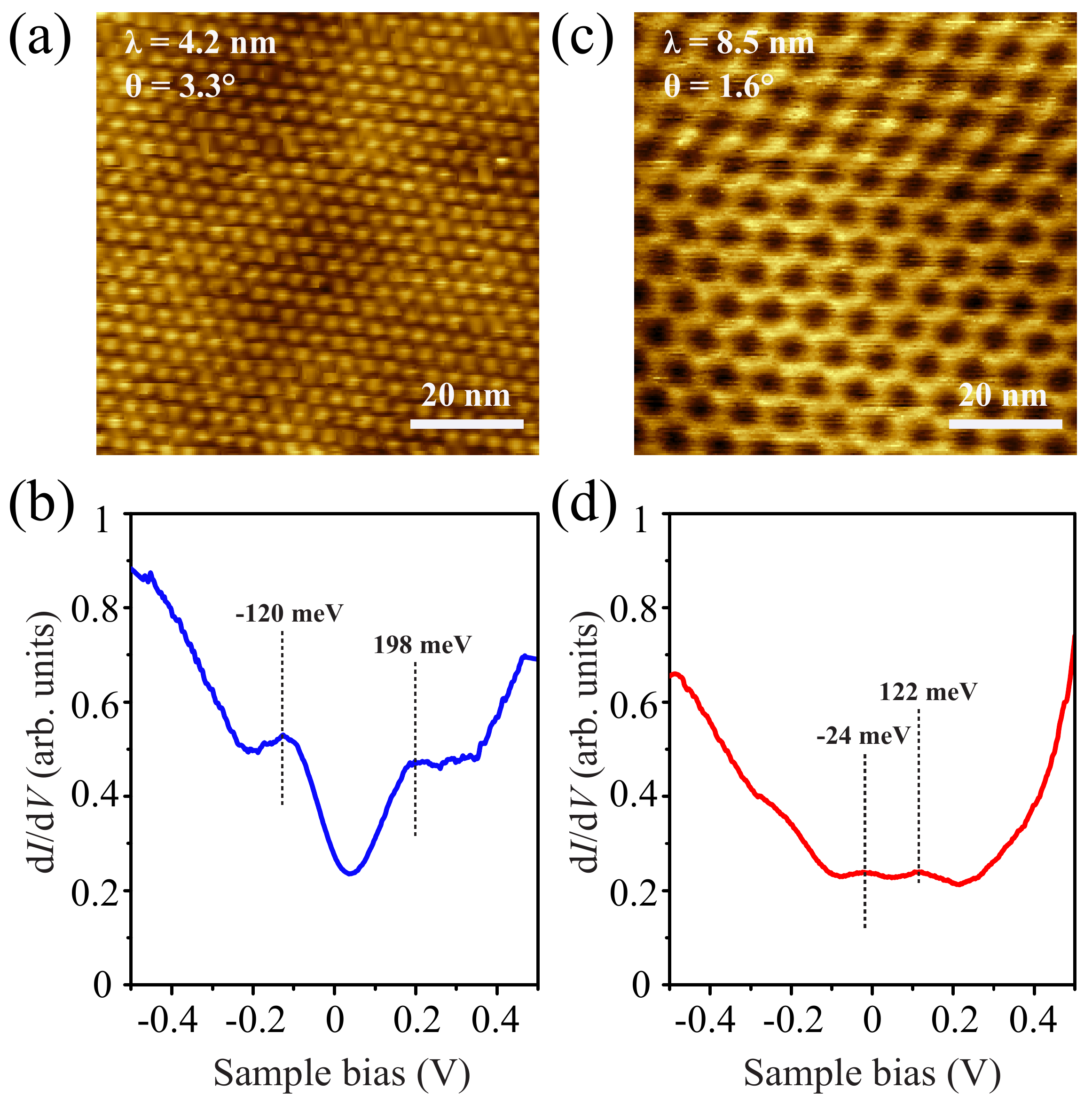}
	\caption{(a) Scanning tunneling microscopy image of twisted bilayer graphene on a h-BN substrate with a twist angle of 3.3\textdegree{}, (b) differential conductivity of the sample with a twist angle of 3.3\textdegree{}, (c) Scanning tunneling microscopy image of twisted bilayer graphene on a h-BN substrate with a twist angle of 1.6\textdegree{} and (d) differential conductivity of the sample with a twist angle of 1.6\textdegree{}. (a) and (c) set points sample bias 400 mV and tunnel current 500 pA. (b) and (d) set points: sample bias -500 mV and tunnel current 200 pA (Reproduced from Q. Yao \textit{et al}., Appl. Phys. Lett. \textbf{116}, 011602 (2020) with permission from American Institute of Physics).}
		\label{fig:moire}
\end{figure}

Of particular interest are the angles at which the electronic bands near the Fermi level become flat, i.e. $v_\text{F} = \hbar^{-1} |\nabla_k E_k| = 0$. The magic twist angles of twisted bilayer graphene have been calculated by Bistritzer and MacDonald \cite{Bistritzer2011}. These authors found that the renormalized Dirac-point band velocity vanishes at twist angles of $\theta_\text{magic} \approx$ 1.05\textdegree{}, 0.5\textdegree{}, 0.35\textdegree{}, 0.24\textdegree{} and 0.2\textdegree{}, respectively \cite{Bistritzer2011}. The reason for the occurrence of a series of magic angles, rather than a single magic angle is due to oscillatory variation of the interlayer hopping energy with the twist angle. The flattening of the energy bands near the Fermi level at these magic angles can be understood qualitatively by comparing the hybridization energy with the kinetic energy. For small angles the energy gap between the two Van Hove singularities is given by,
\begin{equation}
	\Delta E_\text{VHS} = 2 \hbar v_\text{F} K \sin (\theta/2) - 2 w
\end{equation}
where $K = 4\pi/3a$ and $w \approx 0.4t_{\perp} \approx 0.1$ eV the hybridization energy.\\

In the case of AB stacked bilayer graphene an electric field not only results in a shift of the Fermi level, but also to the opening of a band gap. Owing to the broken sub-lattice symmetry charge is shifted from one sub-lattice to the other sub-lattice, which leads to the opening of a bandgap \cite{Oostinga2007,Zhang2009}. The size of this bandgap ($\Delta$) is given by,
\begin{equation}
	\Delta = \sqrt{\frac{e^2V^2t_{\perp}^2}{t_{\perp}^2 + e^2V^2}}
\end{equation}
where $V$ is the applied electrostatic potential across the two graphene layers, $t_{\perp}$ the interlayer hopping energy and $e$ the elementary charge.\\

The same gap opening does, of course, also occur in twisted bilayer graphene. Via gating the Van Hove singularities can be shifted up- or downwards. The maximal number of electrons that can be hosted in the lowest energy bands per unit cell is 8 (2 valleys and 2 spins, both in 2 layers). Electronic instabilities due to correlation effects lead in the vast majority of cases to new physical phenomena, such as magnetic ordering, metal insulator transitions, superconductivity, topological transport, charge or spin density waves, etc. Electronic instabilities often occur when an electronic state with an appreciable density of states, e.g. a Van Hove singularity, crosses the Fermi level. Although the Fermi level can be shifted via doping or gating, the Van Hove singularities of most materials are often located too far from the Fermi level to observe this rich physics. The Van Hove singularities of twisted bilayer graphene can be brought arbitrarily close to the Fermi level and therefore this material is perfectly suited for a systematic study of correlated electron phenomena.
\newpage

\begin{figure}[H] 
	\centering

	\vspace{1cm}
	\includegraphics[trim=0cm 0cm 0cm 0cm,clip=true,width=0.4\textwidth]{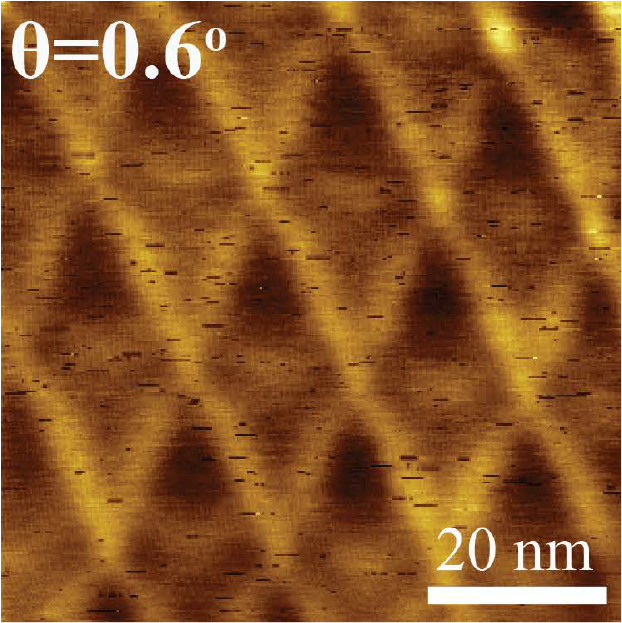}
	\caption{Scanning tunneling microscopy image of twisted bilayer graphene with a twist angle of 0.6\textdegree{}. The hexagonal network is due to valley-projected topologically protected one-dimensional electronic states. Sample bias 400 mV and tunnel current 500 pA. (Reproduced from Q. Yao \textit{et al}., Appl. Phys. Lett. \textbf{116}, 011602 (2020) with permission from American Institute of Physics).}
		\label{fig:network}
\end{figure}

\begin{figure}[H]
	\centering

	\includegraphics[trim=0cm 0cm 0cm 0cm,clip=true,width=\textwidth]{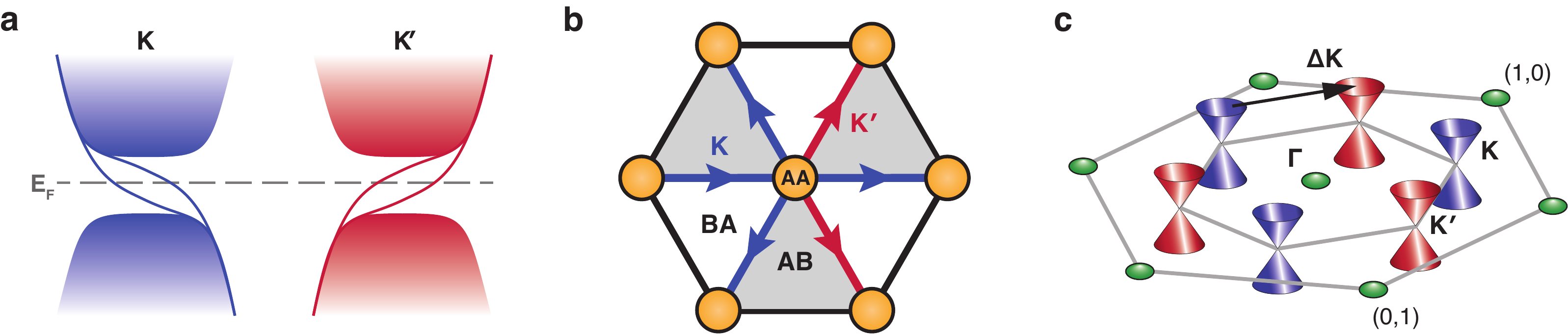}
	\caption{Two-dimensional hexagonal network of topologically protected states in small-angle twisted bilayer graphene. (a) Schematic diagram of the electronic band structure of twisted bilayer graphene upon the application of a transversal electric field at the $K$ and $K’$ valleys, respectively. Two bands cross the Fermi level in each valley. (b) Real-space representation of valley-protected transport channels across the AB/BA domain boundaries, showing the possible scattering directions for a $K$-valley electron arriving at an AA node in the network. The red ($K’$) directions are forbidden. (c) Schematic diagram of graphene in reciprocal space. The (1$\times$1) reciprocal lattice points and the Dirac cones at the $K$ and $K’$ points are labeled by green circles and red/blue cones, respectively. Inter-valley scattering is indicated by the black arrow. (reproduced from J.D. Verbakel  \textit{et al}., Phys. Rev. B \textbf{103}, 165134 (2021) with permission from American Physical Society).}
		\label{fig:intervalley}
\end{figure}

There is another very important difference between twisted bilayer graphene and normal AB stacked bilayer graphene. AB stacked bilayer graphene is single domain regarding the stacking order, whereas twisted bilayer graphene has AB as well as BA domains. These AB and BA domains form an ordered hexagonal structure. As the bandgaps that are opened by a transversal electric field in AB and BA stacked bilayer graphene are ‘inverted’ topologically protected states will emerge at the boundaries between the AB and BA domains. This effect is sometimes also referred to as the quantum valley Hall effect. We have to emphasize here that these topologically protected states can be measured experimentally only when the AB and BA stacked domains are sufficiently large. At twist angles smaller than $\sim$1\textdegree{}, the AB and BA domains are large enough to fulfil this requirement (see Figure \ref{fig:network}). The Berry curvature of the valence band of bilayer graphene is peaked around the two valleys and the two valleys have opposite Chern numbers, resulting in a total Chern number of 0 for the full Brillouin zone. For small biases $V$ ($V$ smaller than the interlayer coupling) the Chern numbers per valley are approximately quantized, i.e. $C_K = -C_{K'} \approx V/|V|$. In case that the two valleys are kept decoupled by any perturbation (i.e. no intervalley scattering), these valley Chern numbers can play the role of a proper topological charge \cite{Yao2009,Zhang2013,Zhang2011,SanJose2013}. If $V$ or the stacking order changes sign, two topologically protected modes per valley will emerge along the boundary between the two domains. Please note that BA and AB stacked regions have opposite Chern valley numbers, i.e. $C_K^{\text{BA}} = -C_K^{\text{AB}}$. It should be noted that we consider here only valley-preserving perturbations. Since the valley Chern numbers change from +1 to -1 and from -1 to +1 of the $K$ and $K’$ valleys, respectively, the total number of conducting channels is 4 (2 channels per valley) resulting a conductance of $4e^2/h$. When a vertical electric field is applied, charge is shifted from one triangular sub-lattice of graphene to the other triangular sub-lattice resulting in the opening of a band gap (see Figure \ref{fig:intervalley}). 

\newpage
If the stacking order is changed from AB (BA) to BA (AB) or the direction of the electric field is changed, the charge shift also changes sign. Intervalley scattering, i.e. scattering from valley to valley, destroys the topological protection. \\

The idea to exploit these valley projected channels that propagate along AB/BA boundaries of bilayer graphene has been pursued by a few groups. By using a split gate in order to produce opposite transverse electric fields Martin \textit{et al}.\cite{Martin2008} were the first to demonstrate that a one-dimensional conducting channel in conventional Bernal stacked bilayer graphene can indeed be realized. The experimental realization of a split gate is, however, quite challenging. Alternatively one could use a single gate and try to produce or find regions with different stacking order. Ju \textit{et al}. \cite{Ju2015} pursued this method and were able to find a few of these one-dimensional valley protected conducting channels. \\

During the last couple of years several groups have studied two-dimensional hexagonal networks of topologically protected states in small-angle twisted bilayer graphene. Huang \textit{et al}. \cite{Huang2018} used low-temperature scanning tunneling microscopy and spectroscopy to image and study the two-dimensional network of topologically protected states.  In another study Rickhaus \textit{et al}. \cite{Rickhaus2018} performed magneto transport measurements. They observed robust Fabry-Pérot and Aharanov-Bohm oscillations in the magnetic field range from 0 T up to 8 T. Their results provide strong evidence that the charge carriers flow in one-dimensional topologically protected channels. Yao \textit{et al}. \cite{Yao2020} used scanning tunneling microscopy to show that the network of topologically protected channels also persists to temperatures as high as room temperature. 

\begin{figure}[H]
	\centering

	\includegraphics[trim=0cm 0cm 0cm 0cm,clip=true,width=0.8\textwidth]{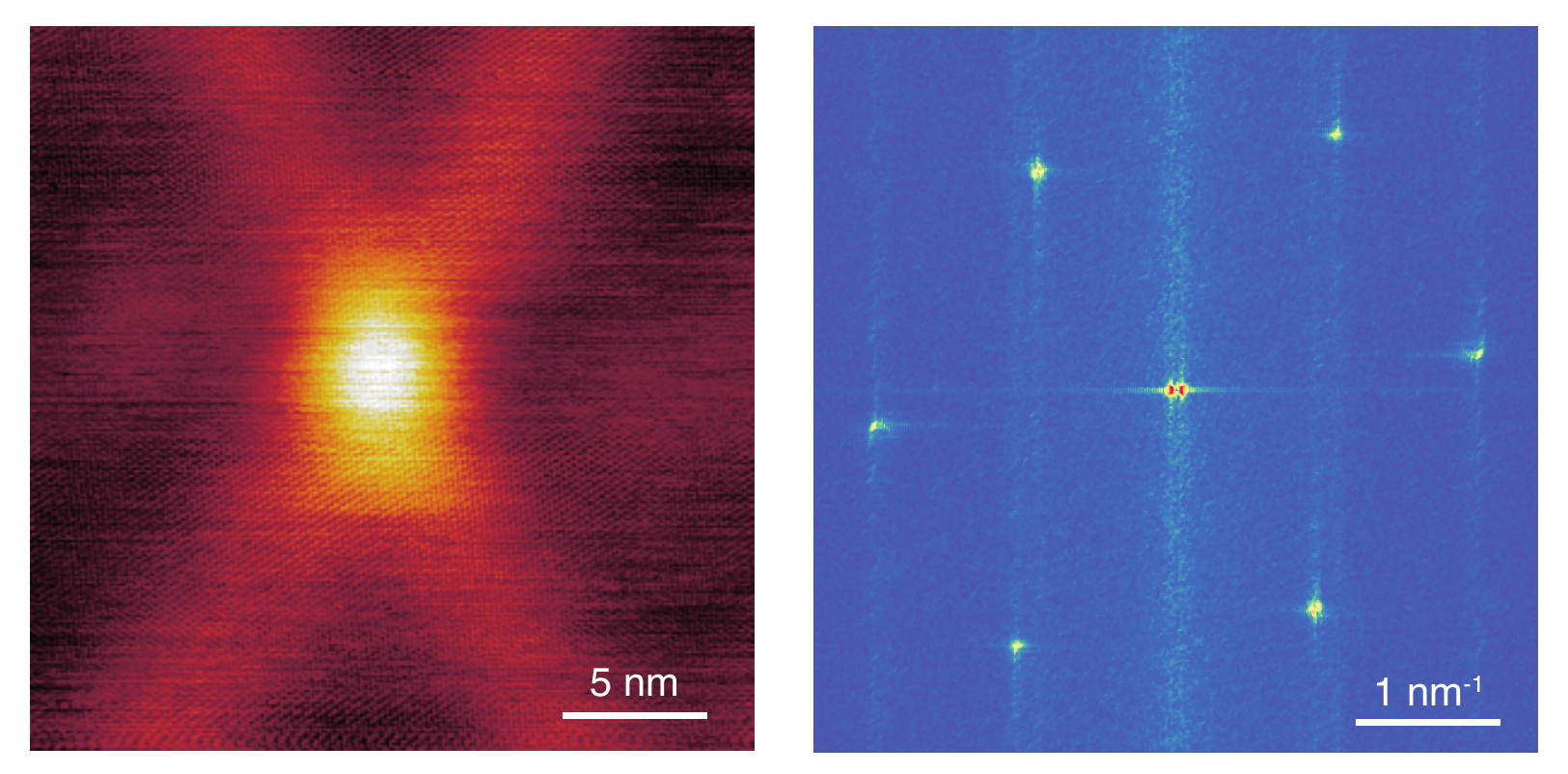}
	\caption{Atomic resolution STM image of 0.6\textdegree{} twisted bilayer graphene and its fast Fourier transform. In the FFT image only the (1$\times$1) spots are visible. The absence of the ($\surd3\times\surd$3)R30\textdegree{} spots demonstrates that scattering from valley to valley does not occur here. (reproduced from J.D. Verbakel  \textit{et al}., Phys. Rev. B \textbf{103}, 165134 (2021) with permission from American Physical Society).}
		\label{fig:fft}
\end{figure}

\newpage
Before proceeding it is important to devote a few words to scattering processes that can occur in single and bilayer graphene. A prerequisite for the occurrence of scattering is the presence of defects or impurities. In case of single layer graphene only scattering from valley to valley, i.e. intervalley scattering, can occur \cite{Brihuega2008,Mallet2012}. Scattering within a valley, also referred to as intravalley scattering, is strongly suppressed in single layer graphene because the pseudospin is locked to the momentum \cite{Yankowitz2012,Luican2011}. This is different for bilayer graphene, where both intravalley as well as intervalley scattering processes can take place. Information regarding scattering processes can be extracted from quasi-particle interference measurements. The interference of incoming and reflected electron waves result in a decaying standing wave pattern in the electron density, also referred to as Friedel oscillations, with a wavevector $2k_\text{F}$. By taking fast Fourier transforms from the differential conductivity ($dI/dV$) or low-bias scanning tunneling microscopy images one obtains valuable information on the various scattering processes. Intravalley scattering results in a circle ($dI/dV$) or disk (topography) with a radius $2k_\text{F}$ that is centered around the $\Gamma$ point. Intervalley scattering (see Figure \ref{fig:intervalley}(c)) results in ($\surd3\times\surd$3)R30\textdegree{} spots in the fast Fourier transform. Verbakel \textit{et al}. \cite{Verbakel2021} used this approach to demonstrate that the electronic transport in the counter-propagating one-dimensional states in small-angle twisted bilayer graphene is indeed valley protected. In Figure \ref{fig:fft} an STM image and corresponding fast Fourier transform of an AA stacked region of a small-angle twisted bilayer graphene hosting one-dimensional topologically protected states are shown. In the fast Fourier transform pattern the ($\surd3\times\surd$3)R30\textdegree{} spots, which are characteristic for intervalley scattering, are missing revealing that the one-dimensional electronic states are indeed topologically protected.

At an AA node, where the one-dimensional channels from the different directions meet, the K and K’ modes can only propagate straight through or make a turn of +60\textdegree{} or -60\textdegree{} as shown in Figure \ref{fig:intervalley} (b). Please note that the scanning tunneling microscopy image is recorded at a sample bias that is located in the bandgap of the bilayer graphene. 

\begin{figure}[H]
	\centering

	\includegraphics[trim=6cm 2cm 6cm 0cm,clip=true,width=0.8\textwidth]{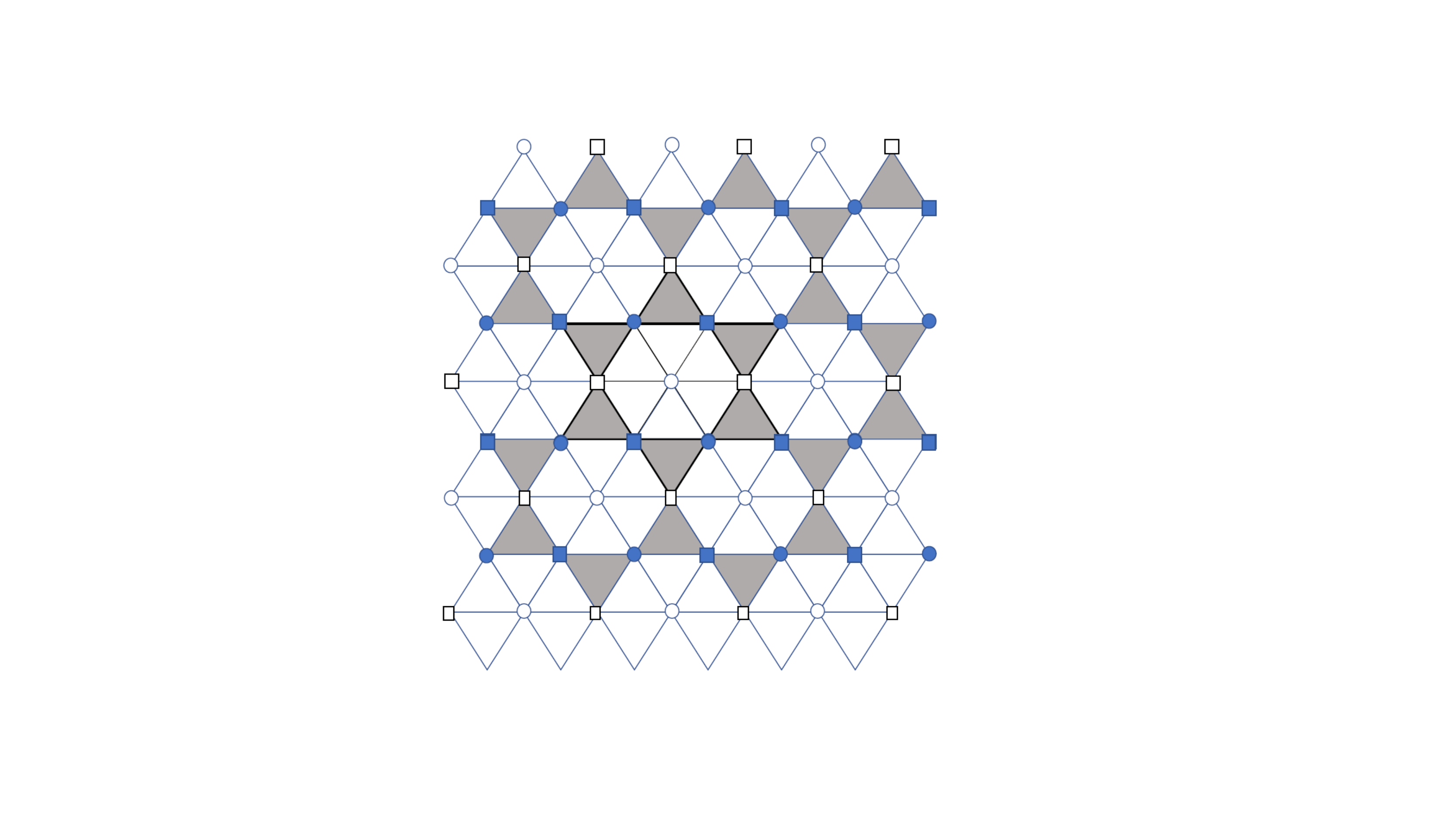}
	\caption{Electronic kagome lattice realized by a twisted bilayer of a two-dimensional material with a buckled honeycomb lattice. The regions with the strongest interlayer coupling are highlighted in grey. The open and filled circles and squares refer to the four AA stacked atom configurations (up-up, down-down, down-up and up-down). Open circles: up-down, open squares: up-up, filled circles: down-down and filled squares: down-up.}
		\label{fig:kagome}
\end{figure}

Twisted bilayers of two-dimensional materials with a buckled honeycomb lattice are more complicated than twisted bilayers graphene. The buckling makes that AA, BB, AB and BA stacked domains occur in four flavors: up-up, down-up, down-up and down-down, respectively (see Figure \ref{fig:kagome}). As the interlayer hopping energy depends on the distance between the atoms an additional modulation of interlayer hopping energy emerges. The down-up configuration has the largest interlayer hopping energy and the up-down configuration the smallest interlayer hopping energy. The up-up and down-down configurations are degenerate and have an interlayer hopping energy that is in between the interlayer hopping energies of the up-down and down-up configurations. Owing to this spatial varying interlayer hopping energies the moiré lattice is electronically modulated, resulting in an electronic kagome lattice. Kagome lattices rarely occur in Nature. There are only a few minerals that have a kagome structure.  The possibility to realize an electronic kagome is of great scientific relevance, particularly for the study of frustrated magnetism and spin liquids. A couple of years ago Li \textit{et al}. \cite{Li2016,Li2018} have realized an electronic kagome lattice by growing multilayer silicene on a Ag(111) substrate. The electrons are localized in the kagome lattice due to destructive interference of the wave functions. The latter leads to a quenching of the kinetic energy and flat energy bands. Li \textit{et al}. \cite{Li2018} found a robust and pronounced one-dimensional edge state that resides at the edges of the kagome lattice.

\newpage
\section{Summary and outlook}
In this report we have given a brief overview of the occurrence of one-dimensional topologically protected electronic states in two-dimensional group IV Dirac materials. We have discussed the quantum spin Hall effect, which relies on the spin-orbit coupling in the materials and gives rise to the development of two counter-propagating topologically protected edge states. These states are spin-polarized because the spin is locked to the momentum. We also show that the application of a transversal electric field in small-angle twisted bilayers of two-dimensional Dirac materials can result in the formation of a two-dimensional network of one-dimensional topologically protected electronic states. These states are, however, only weakly topologically protected. The occurrence of defects or impurities in the twisted bilayer can give lead to intervalley scattering processes that have a detrimental effect on the aforementioned topologically protected states. 

\section*{Acknowledgements}
This work was part of the research program on 2D semiconductor crystals with Project No. FV157-TWOD, which was financed by the Netherlands Organization for Scientific Research (NWO). Q.Y. thanks the China Scholarship Council for financial support.

\newpage
\bibliographystyle{ieeetr}
\bibliography{bibliography.bib}

\end{document}